\documentclass[aps,prd,superscriptaddress,amsmath,amssymb,
nofootinbib]{revtex4} 

\usepackage{natbib}

\usepackage{graphicx,amssymb, amsmath}
\usepackage{dcolumn}
\usepackage{bm}

\newcommand{\gray}{$\gamma$-ray}
\newcommand{\fermi}{{\emph{Fermi}}}
\newcommand{\sas}{{\emph{SAS-2}}}
\newcommand{\oso}{{\emph{OSO-3}}}

\begin{document}


\title{\fermi{} Large Area Telescope Observations of the Cosmic-Ray Induced
\gray{} Emission of the Earth's Atmosphere}




\author{A.~A.~Abdo}
\affiliation{Space Science Division, Naval Research Laboratory, Washington, DC 20375, USA}
\affiliation{National Research Council Research Associate, National Academy of Sciences, Washington, DC 20001, USA}
\author{M.~Ackermann}
\email{markusa@slac.stanford.edu}
\affiliation{W. W. Hansen Experimental Physics Laboratory, Kavli Institute for Particle Astrophysics and Cosmology, Department of Physics and SLAC National Accelerator Laboratory, Stanford University, Stanford, CA 94305, USA}
\author{M.~Ajello}
\affiliation{W. W. Hansen Experimental Physics Laboratory, Kavli Institute for Particle Astrophysics and Cosmology, Department of Physics and SLAC National Accelerator Laboratory, Stanford University, Stanford, CA 94305, USA}
\author{W.~B.~Atwood}
\affiliation{Santa Cruz Institute for Particle Physics, Department of Physics and Department of Astronomy and Astrophysics, University of California at Santa Cruz, Santa Cruz, CA 95064, USA}
\author{L.~Baldini}
\affiliation{Istituto Nazionale di Fisica Nucleare, Sezione di Pisa, I-56127 Pisa, Italy}
\author{J.~Ballet}
\affiliation{Laboratoire AIM, CEA-IRFU/CNRS/Universit\'e Paris Diderot, Service d'Astrophysique, CEA Saclay, 91191 Gif sur Yvette, France}
\author{G.~Barbiellini}
\affiliation{Istituto Nazionale di Fisica Nucleare, Sezione di Trieste, I-34127 Trieste, Italy}
\affiliation{Dipartimento di Fisica, Universit\`a di Trieste, I-34127 Trieste, Italy}
\author{D.~Bastieri}
\affiliation{Istituto Nazionale di Fisica Nucleare, Sezione di Padova, I-35131 Padova, Italy}
\affiliation{Dipartimento di Fisica ``G. Galilei", Universit\`a di Padova, I-35131 Padova, Italy}
\author{B.~M.~Baughman}
\affiliation{Department of Physics, Center for Cosmology and Astro-Particle Physics, The Ohio State University, Columbus, OH 43210, USA}
\author{K.~Bechtol}
\affiliation{W. W. Hansen Experimental Physics Laboratory, Kavli Institute for Particle Astrophysics and Cosmology, Department of Physics and SLAC National Accelerator Laboratory, Stanford University, Stanford, CA 94305, USA}
% ...


\collaboration{Fermi-LAT collaboration}

\date{\today}

\begin{abstract}
  We report on measurements of the cosmic-ray induced \gray{} emission
  of Earth's atmosphere by the Large Area Telescope onboard the
  \fermi{} Gamma-ray Space Telescope.  The LAT has observed the Earth
  during its commissioning phase and with a dedicated Earth-limb
  following observation in September 2008.  These measurements yielded
  $\sim 6.4 \times 10^{6}$ photons with energies $>100$ MeV and $\sim
  250$ hours total livetime for the highest quality data selection.
  This allows the study of the spatial and spectral distributions of
  these photons with unprecedented detail.  The spectrum of the
  emission -- often referred to as Earth albedo gamma-ray emission --
  has a power-law shape up to 500 GeV with spectral index
  $\Gamma = 2.79 \pm 0.06$.
\end{abstract}


\maketitle

\section{Introduction}

For an earth-orbiting \gray{} detector such as the Large Area
Telescope (LAT) on board the Fermi \gray{} Space Telescope
({\emph{Fermi}}), the Earth is the brightest source in the sky due to
its proximity.  The \gray{} emission from the Earth is produced by
cosmic-ray interactions with the Earth's atmosphere, and is often
called the \gray{} albedo.  The vast majority of cosmic rays are
protons and heavier nuclei.  Atmospheric \gray{} emission is the
result of cosmic-ray cascades, mainly through the decay of neutral
pions and kaons and through Bremsstrahlung of electrons and positrons
(dominating at energies below $\sim$50 MeV \citep{Morris1984}).  Due
to the kinematics of the collisions, the cross sections of these
processes at high energies are peaked in the forward direction.

The spectrum of the albedo \gray{s} is not uniform across the Earth's
disk.  The spectrum of the inner part of the disk is soft owing to the
small number of secondary particles, typically neutral pions, that are
back-scattered at large angles relative to the direction of the
cosmic-ray cascade developing deep into the atmosphere.  Cosmic rays
that enter the atmosphere near grazing incidence produce showers whose
forward-moving \gray{s} can penetrate the thin atmospheric layer
making the limb bright when viewed from orbit.
The geometric effect of shower production and absorption in the
atmosphere producing a bright \gray{} horizon was first calculated by
\citet{Stecker1973}.  A Monte Carlo model of \gray{} production by
cosmic-ray interactions in the atmosphere later developed by
\citet{Morris1984} produced reasonable agreement with the spatial and
spectral measurements by various balloon-borne and spacecraft
instruments.

The Earth's \gray{} emission has been observed by several previous
satellite missions starting as early as the 1960s with a detector
flown on the third Orbiting Solar Observatory, \oso{}, for energies
above 50~MeV~\citep{OSO3Earth}.  This instrument detected an
enhancement of \gray{s} when pointing towards the Earth, consistent
with balloon flight measurements~\citep{fichtel1969, kinzer1974,
  fishman1976, ryan1977, schoenfelder1977} and followed up by other
satellite missions~\citep{golenetskiy1975, imhof1976, guryan1979,
  Share2001}.  A more detailed study of the \gray{} emission from
interactions of cosmic rays with the Earth's atmosphere was done by
~\citet{SASEarth} using approximately 6700 \gray{} events from the
Earth, recorded with the second Small Astronomy Satellite, \sas{}.
These early measurements showed a clear peak of \gray{} intensity
towards the Earth horizon, with a factor of $\sim$10 larger intensity
than seen towards the nadir direction. Also they show a modulation of
the \gray{} intensity with respect to the east-west direction as
expected from the deflection of cosmic rays in the Earth's magnetic
field.  These findings were confirmed and extended to higher energies
with the high-energy \gray{} satellite detector on-board the
\emph{Compton Gamma-Ray Observatory (CGRO)}, the Energetic \gray{}
Experiment Telescope (EGRET).  EGRET was in operation from 1991 to
2000, and approximately 60\% of the mission's total 5.2 million events
were measured from the Earth direction~\citep{PetryEarth}.

The next generation \gray{} observatory, \fermi{}, was launched on 11
June 2008.  The primary instrument on \fermi{}, the Large Area
Telescope (LAT), is a pair-conversion telescope that detects \gray{s}
through conversion into an e$^+$-e$^-$-pair.  The trajectories of this
pair are recorded in the tracker and allows for the reconstruction of
the direction of the incident photon.  The subsequent particle shower
development is sampled in the tracker and a calorimeter to determine
the photon energy.  The LAT's wide field of view ($\sim 2.4$~sr),
effective area ($\sim 8000$ cm$^2$ at 1 GeV), improved point spread
function (PSF), and broad energy coverage (20 MeV to $>300$ GeV)
provide significantly improved sensitivity over its
predecessors~\citep{LATPaper}.

During routine science operations (the so-called ``survey'' mode), the
LAT points away from the Earth because of the relative brightness of
the Earth at GeV \gray{} energies and the background it would
introduce to astrophysical sources.  The center of the field-of-view
($z$-axis) of the LAT is kept within $35^{\circ}$ relative to the
zenith (the direction pointing away from the center of the Earth) for
most of the time \footnote{This rocking angle has recently been
  increased to $50^{\circ}$}. However, during the commissioning phase
of the instrument, the pointing was such that parts of the Earth were
within the field-of-view at varying angles with respect to the LAT
$z$-axis.  In addition, during September 2008 a dedicated observation
of the Earth's limb was made.  This combined dataset can be used for
the analysis of the albedo emission.

There are several reasons that render a study of the Earth's \gray{}
albedo with the LAT useful.  First, measurements of the albedo
\gray{s} contribute to the understanding of the celestial \gray{}
background for the LAT and future satellite instruments.  Second, such
observations can be used for a calibration of the LAT instrument.
Finally, if the production of secondary \gray{s} in the Earth's
atmosphere is sufficiently well understood and modeled, observations
of the albedo emission yield a measurement of the interactions of
primary cosmic rays with the Earth's atmosphere and can, therefore,
provide information on the spectrum of primary cosmic rays,
interaction cross sections, atmospheric properties, and the deflection
of primary cosmic rays by the geomagnetic field.

The current work describes the analysis of the Earth's \gray{s} albedo
emission at energies above 100 MeV observed with the LAT.

\section{Dataset}

Two main data sets have been used: (i) the Launch \& Early Operations
(LEO) data taken during the first 60 days of the mission, during which
time the Earth limb was often closer to the field of view than during
a regular sky survey (only data were used that were suitable for
science observations), (ii) two orbits (i.e.\ close to 3 hours) of
direct limb observations during which the Earth limb was kept within
the field of view with the aim of adding on-axis photons to the
overall data set.  The LEO data contain many different observation
modes, but the main modes are a north-orbital-pole pointing and a
pointing mode with the LAT $z$-axis centered at the Vela pulsar.  All
results shown here are averaged over many orbits of \fermi{} and
correspond to the highest quality event selection currently available
for the LAT data (the so-called ``diffuse'' class events
\citep{LATPaper}).  The total number of events $> 100$~MeV in our
dataset is $\sim 6.4 \times 10^{6}$ corresponding to $\sim 250$ hours
of livetime.  The maximum total exposure is close to the Earth north
pole with $\sim 2 \times 10^{9}$ cm$^{2}$ s at 1 GeV.

In addition to the standard diffuse cuts, corresponding to the
post-launch instrument response functions {\emph{P6V3}}, we applied
the following selections:
\begin{itemize}
\item $\Theta_{\mathrm{nadir}} < 80^{\circ}$ to select photons
  coming from the earth.
\item $\Theta_{\mathrm{LAT}} < 65^{\circ}$ to avoid the edge of the
  field of view of the LAT ($\Theta_{\mathrm{LAT}}$ being the angle with
  respect to the LAT z-axis). 
\end{itemize}

\section{Results}

\subsection{Intensity Maps}
To generate intensity maps we transform the incident photon directions
into the Earth-centric coordinate system with angles
$\Theta_{\mathrm{nadir}}$ and $\Phi_{\mathrm{nadir}}$ denoting the
zenithal and azimuthal angle about the direction pointing from the
spacecraft towards the center of the Earth, where
$\Phi_{\mathrm{nadir}} = 0$ corresponds to the north.  The photons
were binned in solid angle for energy intervals with spacing 10 bins
per decade from 200 MeV up to 500 GeV.  The solid angle bins were
spaced equally in $5^\circ$ bins in $\Phi_{\mathrm{nadir}}$ with an
energy dependent binning in $\Theta_{\mathrm{nadir}}$ changing
from $1^\circ$ at 200 MeV to $0.1^\circ$ above 10~GeV to take into
account the LAT's energy-dependent angular resolution.  The LAT
pointing and livetime history were evaluated in the same coordinate
system in 30-second time steps, creating maps of the exposure of the
spacecraft to allow conversion from photon counts per solid angle to
intensities.

\begin{figure}
  \centering
  \includegraphics[width=\textwidth]{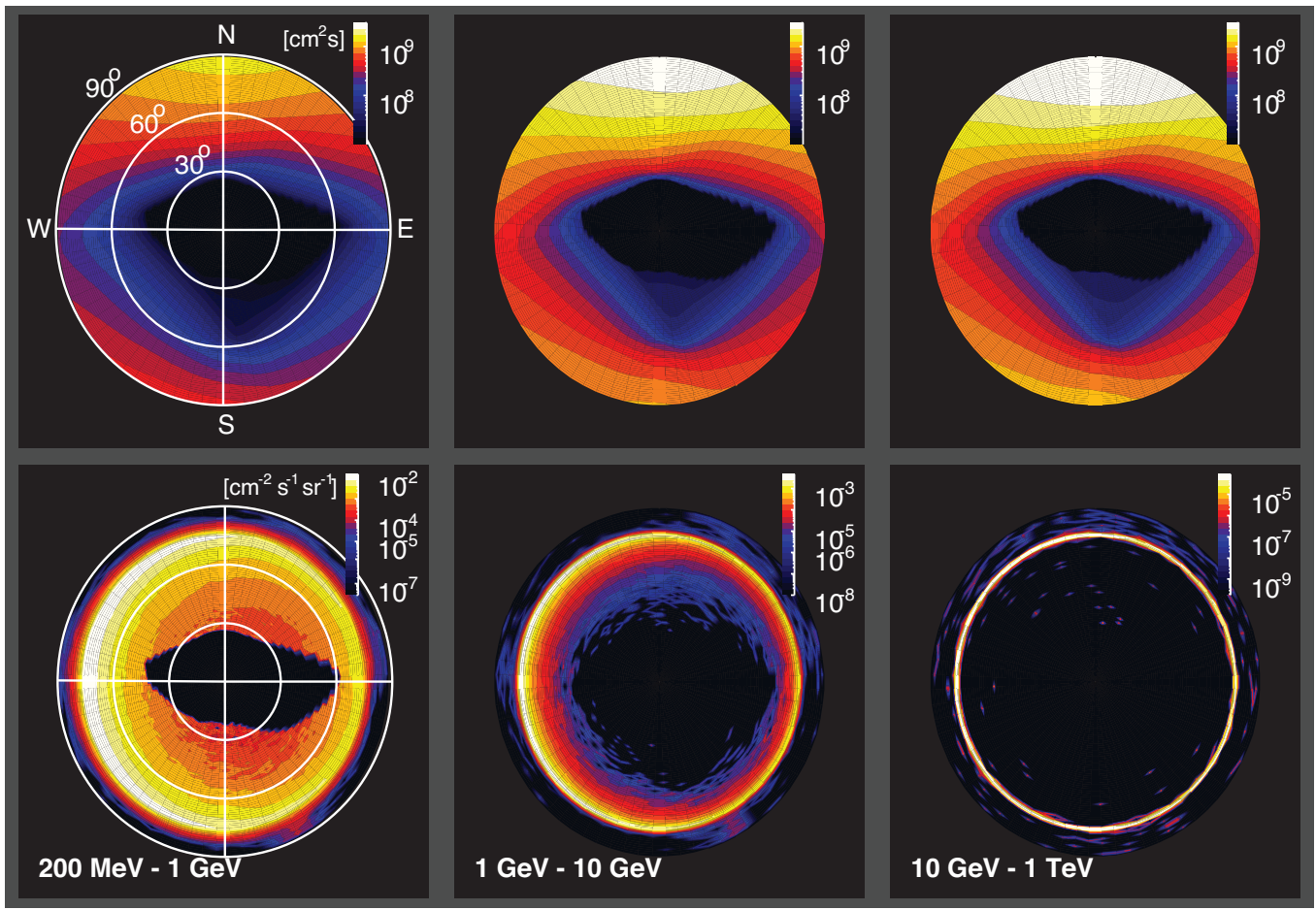}
  \caption{Two-dimensional maps of exposure (top row) and intensity
    (bottom row) for three different energy bands. The coordinate
    system in these maps is a polar representation such that the
    nadir-direction (i.e. looking down towards the center of the
    Earth) with $\Theta_{\mathrm{nadir}} = 0^{\circ}$ is at the center
    of the plot. $\Theta_{\mathrm{nadir}}$ is increasing in radial
    direction with the edge of the plot towards the horizon at
    $\Theta_{\mathrm{nadir}} = 90^{\circ}$. The north pole is to the
    top and west is to the left. The exposure maps are given in units
    of cm$^2$ s, the intensity maps in units of cm$^{-2}$ s$^{-1}$
    sr$^{-1}$.
    \label{fig:Maps}}
\end{figure}

Figure~\ref{fig:Maps} shows the resulting exposure and photon
intensity maps for 3 different energy bands (left: 200~MeV to 1~GeV,
center: 1~GeV to 10~GeV, right: 10~GeV to 500~GeV).  The center of
each plot (i.e.\ $\Theta_{\mathrm{nadir}} = 0^{\circ}$) is the nadir
direction, while the edge of the plots are the horizon at LAT's
altitude ($\Theta_{\mathrm{nadir}} = 90^{\circ}$).  North is to the
top of the maps and west to the left.  All intensity maps have been
background corrected by subtracting the average intensity from a ring
between $\Theta_{\mathrm{nadir}} = 79^{\circ}$ to $80^{\circ}$
averaged over $\Phi_{\mathrm{nadir}}$.

The exposure is a factor $\sim 2$ higher in the north compared to
south (top row).  This is due to the observation strategy in which the
majority of the data were taken during a north-orbital pole pointing.
The exposure increases with energy due to the increase in the LAT
effective area.  It can also be seen that towards the nadir the
exposure is much lower than towards the rim.  To avoid the
introduction of large errors from low exposure bins in the spectral
analysis, areas where the exposure is less than 0.5\% of the maximum
exposure in the corresponding energy band are excluded from the
analysis in all energy bands.  The discarded areas due to low exposure
match the black regions in the center of the exposure maps.  Compared
with EGRET, the LAT exposure is about a factor of $\sim 30$
larger~\citep{PetryEarth}.

The \gray{} intensity maps (bottom row) in units of photons cm$^{-2}$
s$^{-1}$ sr$^{-1}$ show the bright limb of the Earth, with the
emission becoming less intense toward nadir.  The Earth limb is very
bright with intensities of up to 0.01 photons cm$^{-2}$ s$^{-1}$
sr$^{-1}$ in the low energy bands -- more than one order of magnitude
brighter than the center of the Galaxy for the same energy range.  The
narrowing of the emission with increasing energy is an effect of both
the PSF of the LAT improving at higher energies~\citep{LATPaper} and
the increasing contribution by \gray{s} produced in the forward
direction of the cosmic-ray showering at higher energies.  Only 27
photons above 10~GeV are detected with angles $\Theta_{\mathrm{nadir}}
< 60^{\circ}$.  The maximum of the \gray{} emission is detected from a
narrow region at $\Theta_{\mathrm{nadir}} \sim 68^{\circ}$.  The
intensity maps also show significant modulation between east and west
with the effect decreasing as the \gray{} energy increases (this is
the so-called ``east-west effect'').  Photons arriving from directions
excluded from the analysis due to low exposure are not shown on the
maps.
  
Our measurements are in a good qualitative agreement with
expectations from standard cosmic-ray air-shower phenomenology.

The east-west effect that can be seen in the lower energy intensity
maps is due to the deflection of cosmic rays in the Earth's magnetic
field.  If there was no magnetic field the cosmic ray intensities at
the top of the atmosphere would be isotropic, and no east-west
asymmetry would exist.  However, the presence of the magnetic field
biases the cosmic-ray intensities interacting in the atmosphere.
There are fewer cosmic rays interacting from an easterly direction
than the west because the magnetic field disallows trajectories from
the east.  This effect is energy dependent so that low-energy cosmic
rays are affected more than high-energy cosmic rays (those with
energies above $\sim50$ GeV are unaffected).  Since the majority of
the emission near the limb comes from viewing the development of the
cosmic-ray showers in the forward direction, the reduced cosmic-ray
intensity from the east effectively suppresses the shower products
coming from this direction.  This effect vanishes when the atmospheric
photons are predominantly generated by cosmic rays that have large
enough energies to allow them to reach the atmosphere undeflected by
Earth's magnetic field.  This is the case for secondary photons with
energies $\gtrsim 10$~GeV.

The high-energy \gray{s} ($E_\gamma > 10$~GeV) are concentrated in a
very narrow band toward, but not exactly aligned with, the Earth-limb
angle, which is at $66.7^\circ$ for the LAT's orbital altitude of 565
km.  At these high energies the number of back-scattered \gray{s}
formed in the shower development is negligible and the intensity is
dominated by forward-emitted \gray{s} from incident cosmic rays with
tangential directions to the Earth's surface.  Vertically incident
showers have their first interaction in the range between 10--20~km,
which corresponds to $\Theta_{\mathrm{nadir}} \sim 67^{\circ}$.  The
atmospheric column density in the limb direction is large, being many
attenuation lengths of material, which suppresses the \gray{} emission
close to the limb.  In addition, the LAT views mostly
grazing-incidence cosmic rays in the higher parts of the atmosphere.
Therefore the maximum of the emission is in the range between
$\Theta_{\mathrm{nadir}}= 67^{\circ}$ to $69^{\circ}$.  For zenith
angles larger than about $\Theta_{\mathrm{nadir}} = 70^{\circ}$ the
atmospheric depth left to produce a hadronic shower becomes too low
and, therefore, very few \gray{s} are detected.  For zenith-angles
smaller than the limb angle at $\Theta_{\mathrm{nadir}} \sim
66.7^{\circ}$, the Earth and its atmosphere form an opaque barrier to
forward-emitted \gray{s}.  Only the significantly lower-intensity
back-scattered and off-axis emission from the cosmic-ray shower
development is detected for these viewing angles.  Thus, the maps
above 10~GeV show a very narrow band close to but not at the Earth's
limb.

\subsection{Spectrum}
\begin{figure}
  \centering
  \includegraphics[width=0.8\textwidth]{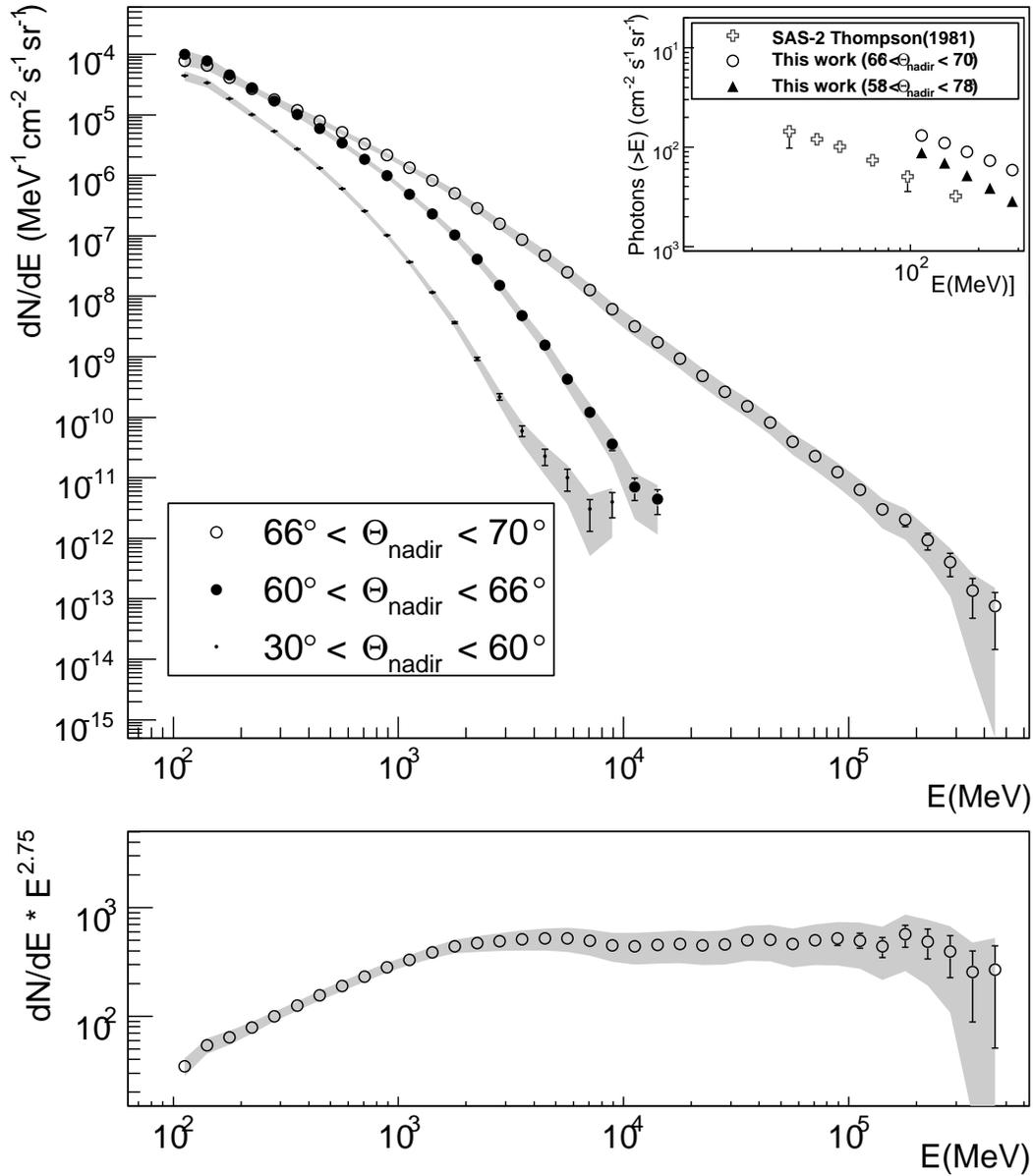}
  \caption{Differential energy spectrum for the \gray{s} produced in
    the Earth's atmosphere taken in different
    $\Theta_{\mathrm{nadir}}$-bands. The spectrum of the limb (open
    circles) shows a power-law behavior. Also shown are \gray{}
    spectra for the inner parts of the Earth's disk (filled circles
    and points). Indicated are both statistical and systematic errors
    (grey band), the latter derived from studies of the Vela
    pulsar. The inset shows a comparison to previous data by the \sas\ 
    satellite~\citep{SASEarth}. To allow for a direction comparison,
    the limb-spectrum is also shown for a larger integration region
    ($58^{\circ} < \Theta_{\mathrm{nadir}} < 78^{\circ} $)
    corresponding to the \sas\ measurement. The bottom panel shows the
    limb-spectrum multiplied by energy to the power 2.75.
\label{fig::Spectrum}}
\end{figure}

Figure~\ref{fig::Spectrum} shows the LAT-measured energy spectrum of
the atmospheric \gray{s} for different nadir bands integrated over
azimuth angles.  The spectra have been corrected for spill-over
effects due to the LAT PSF.  This was done by measuring the intensity
in each energy bin outside of the Earth limb, where very little
emission is expected due to the significantly lower atmospheric
density. Outside of the rim is defined here as $70^{\circ} <
\Theta_{\mathrm{nadir}} < 73^{\circ}$. Only for the low energy part of
the spectrum (below $\sim 1.5$~GeV) this spill-over effect is
relevant. Beyond that energy the intensity outside the rim is lower by
many orders of magnitude than the intensity from inside the rim.

The average differential spectrum of the limb region (open circles),
i.e.\ for $66^{\circ} < \Theta_{\mathrm{nadir}} < 70^{\circ}$, shows
evidence for a turn-over at energies below 3~GeV.  Between 3~GeV and
500~GeV the spectrum can be fitted with a power-law $dN/dE = I_0\ E ^
{-\Gamma}$ with normalization $I_0 = (7.4 \pm 1.0)\times 10^{2}$
MeV$^{-1}$ cm$^{-2}$ s$^{-1}$ sr$^{-1}$ and photon index $\Gamma =
2.79\pm0.06$.  Since both the intensity and the exposure vary
considerably over the limb region, a derivation of the statistical
errors is not straightforward. To determine the statistical errors we
use a Monte Carlo technique: 1000 pseudo-count experiments are
generated for each energy bin, using the LAT exposure and a model of
the gamma-ray intensity emitted by the Earth albedo derived from EGRET
observations~\citep{PetryEarth}. In this model, the average intensity
from the limb is renormalized to match the intensities reported
here. The RMS of the intensities derived from these pseudo-experiments
is then used as the uncertainty of the reported gamma-ray intensity
from the Earth limb.

For the fit, the systematic errors (shown as grey band) have been
taken into account, conservatively adding them linearly to the
statistical errors. These systematic errors have been derived from
Vela observations by comparing the measured \gray{s} to various
simulation of \gray{s} from Vela where these simulations take the
uncertainties in the instrument response functions into
account~\citep{FermiVela}. These systematic errors therefore
  essentially show the contribution from the uncertainties in the
  effective area of the instrument. For energies above 10~GeV, the
systematic errors are less well known (due to the Vela cutoff) and
have been extrapolated from the value at 10~GeV. For this
high-statistics data set, the systematic errors dominate over the
statistical errors nearly up to the highest energies.

The bottom panel of Fig.~\ref{fig::Spectrum} shows the differential
intensity multiplied by $E^{2.75}$ -- the spectral index of cosmic-ray
protons~\citep{ryan1972,BESS}. The $\chi^2$ of the fit is 1.2 for 19
degrees of freedom when using the systematic errors in the fitting
 ($\chi^2$/ndf = 89/19 when using only statistical errors).

The integrated intensity of Earth's limb above 100~MeV is $5.2\times
10^{-3}$ cm$^{-2}$ s$^{-1}$ sr$^{-1}$, which makes the Earth by far
the single brightest source in the LAT energy range.  For this data
set there are 16 photons above 500~GeV and 3 events above 1 TeV with
the highest energy photon having an energy of 1.14 TeV.  Events above
500~GeV have not been used in the spectral analysis since the energy
calibration of the LAT is still under investigation at these energies.
Also shown as solid circles and dots are the spectra for two inner
regions of the Earth's disk ($60^{\circ} < \Theta_{\mathrm{nadir}} <
66^{\circ}$ and $30^{\circ} < \Theta_{\mathrm{nadir}} < 60^{\circ}$).
The spectra become much softer above a few hundred MeV when moving
towards the inner part of the disk.  Fitting the inner parts by a
power-law with energy-dependent index of the form $dN/dE = I_0\ E^
{-\Gamma - \beta log_{10}(E/MeV)}$ yields $\Gamma = 7.76 \pm 3.1$ and
$\beta = 1.78 \pm 0.5$ for the range between $60^{\circ} <
\Theta_{\mathrm{nadir}} < 66^{\circ}$ and $\Gamma = 6.9 \pm 0.1$ and
$\beta = 1.90 \pm 0.02$ for the range between $30^{\circ} <
\Theta_{\mathrm{nadir}} < 660^{\circ}$.

The spectrum of the Earth's limb is dominated by \gray{s} from cosmic
ray interactions in the upper atmospheric layers (in tangential
directions) pointed towards the LAT that do not suffer large energy
losses in the atmosphere (as compared to the back-scattered \gray{s}).
At high \gray{} energies (above 3.2~GeV) the spectrum enters the
regime of cosmic-ray primaries ($\gtrsim 10$~GeV) unaffected by the
Earth's magnetic fields, and, therefore, should have a spectral index
close to that of cosmic rays \citep{Aharonian2000}.
The fitted power-law index of the atmospheric \gray{} emission from
the limb is $\Gamma = 2.79 \pm 0.05$, compared to an index of $\Gamma
= 2.75 \pm 0.03$ for the primary cosmic-ray spectrum.  Since the
spectrum of cosmic rays is well measured, eventually, the \gray{s}
from the Earth's atmosphere might help to calibrate the effective area
of the LAT beyond the energy regime where we have test-beam data. As
previously described, moving from the limb to the inner regions of the
Earth, the LAT moves from measuring the CR showers from a
predominantly forward direction to a side-on view and then to a
backward direction.  This results in a much steeper spectrum for the
inner parts with very few \gray{s} above 1~GeV as shown by the filled
circles.

The inset in the top panel of Fig.~\ref{fig::Spectrum} shows a
comparison with previous results as published in~\citet{SASEarth}. The
spectra are qualitatively similar when taking into account the region
for which the \sas\ spectrum had been derived ($\pm 10^{\circ}$ around
the Earth limb). The differences in the measured flux could be due to
the limited energy resolution of \sas\ or the fact that when the PSF
is folded into the integration region, the match in areas over which
the measurements were taken is not perfect.

\subsection{Azimuthal and Zenithal profiles}

\begin{figure}
  \centering
 \includegraphics[width=\textwidth]{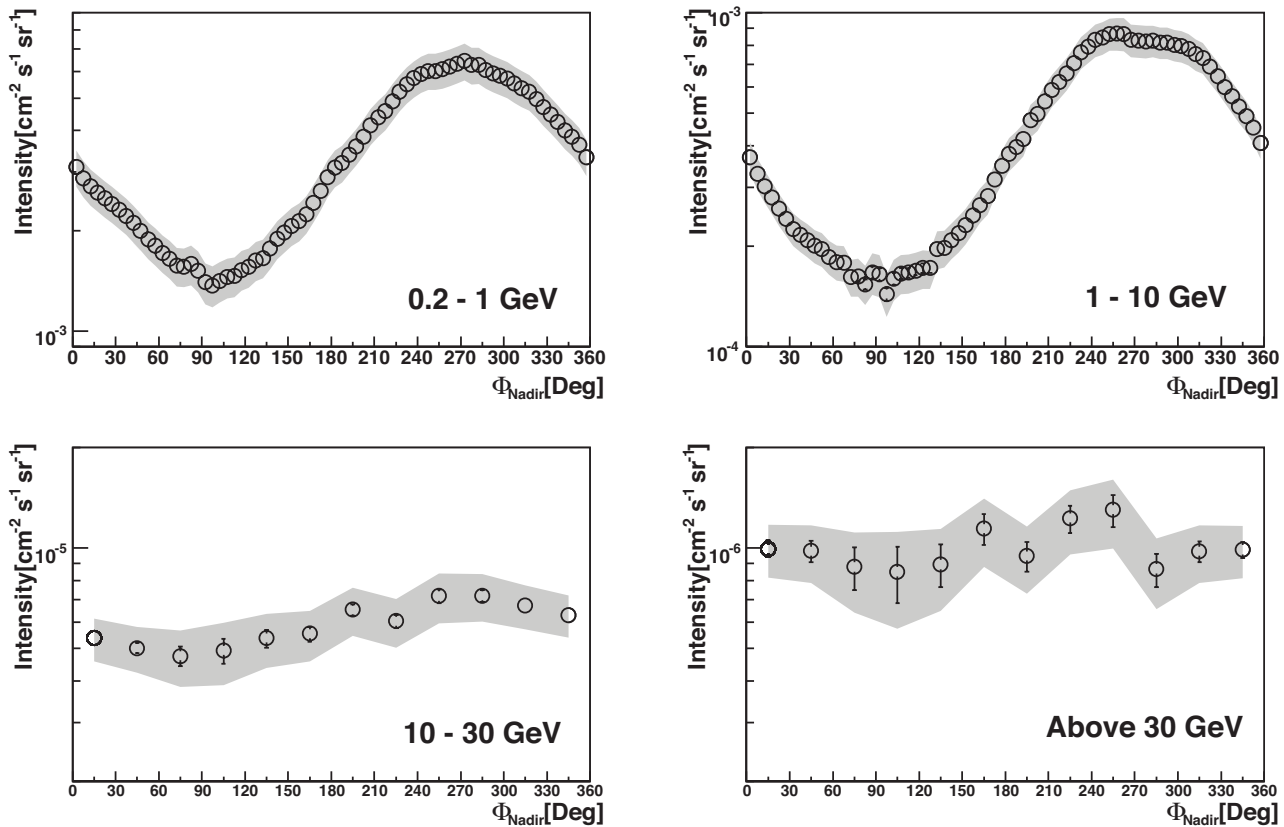}
  \caption{Photon intensity as a function of azimuth for 4 different energy
    bands along the limb ($60^{\circ} < \Theta_{\mathrm{nadir}} <
    75^{\circ}$). Systematic errors as derived from LAT data on the
    Vela pulsar are shown as grey bands. A modulation is
    apparent in the low-energy bands which fades away at higher
    energies (see text for more details). 
    \label{fig::Azimuth}}
\end{figure}

Figure~\ref{fig::Azimuth} shows an azimuthal profile of the limb
emission (i.e. for $60^{\circ} < \Theta_{\mathrm{nadir}} <
75^{\circ}$) for four different energy bands (0.2-1~GeV, 1-10~GeV,
10-30~GeV and above 30~GeV).  As can be seen, the east-west effect is
a very prominent modulation at low energies with a factor of $\sim 5$
intensity difference between the minimum in the east (at
$\Phi_{\mathrm{nadir}} = 90^{\circ}$) and the maximum in the west (at
$\Phi_{\mathrm{nadir}} = 270^{\circ}$).  This factor is consistent
with what has been shown for EGRET data~\citep{PetryEarth}.  The
difference in intensity between the east and the west for the highest
energy band is consistent within errors with no modulation ($(8.0 \pm
2.0) \times 10^{-7} \mathrm{cm}^{-2} \mathrm{s}^{-1} \mathrm{sr}^{-1}$
from the east and $(10.0 \pm 2.0) \times 10^{-7} \mathrm{cm}^{-2}
\mathrm{s}^{-1} \mathrm{sr}^{-1}$ from the west).  The no-modulation
hypothesis provides an acceptable fit only for the $>30$~GeV band
($\chi^2 = 74$ for 71 degrees of freedom).  All other bands show the
variation in intensity with azimuth and hence a poor fit to a constant (for
example, the $10-30$~GeV band has a $\chi^2 = 188$ for 71 degrees of
freedom).  A comparison between the intensity from the north and the
intensity 
from the south is consistent within errors (variation is below 15\%).
The behavior of these azimuthal variations can be qualitatively
understood taking into account the production mechanism.  Low-energy
cosmic-ray shower products are altered by Earth's magnetic field and
due to their predominantly positive charge preferentially selects
particles coming from the west.  Higher-energy cosmic rays have higher
magnetic rigidities and are therefore less affected by Earth's
magnetic field.

\begin{figure}
  \centering
  \includegraphics[width=\textwidth]{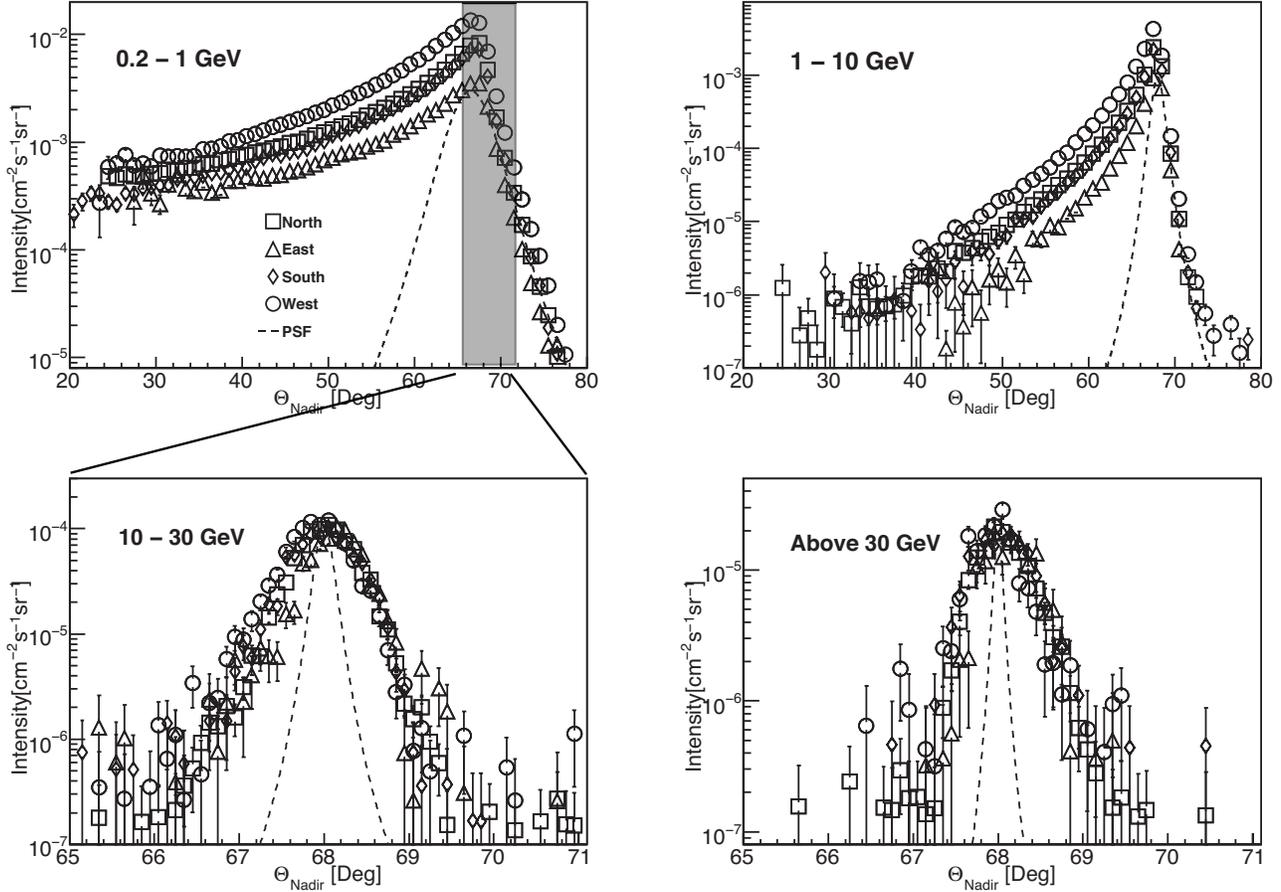}
  \caption{Average intensity as a function of nadir angle in wedges towards the
    different directions (north, west, south, east) for 4 different
    energy bands. The narrowing of the emission with
    increasing energy is seen as well as the change in shape (with a broad tail
    towards the center of the Earth for low energies) and a very
    narrow limb for the high energies. The comparison of the PSF
    calculated for the lower bound of the energy bin (averaged between
    front and back-converting events) is shown as a
    dashed line. Note that the range of the histogram and the binning
    is different for the two high-energy bands.
    \label{fig::Zenith}}
\end{figure}

Fig.~\ref{fig::Zenith} shows the zenith angle distribution for several
energy bands separated into angular segments of $90^{\circ}$
referenced to the nadir centred on the directions north, east, south,
and west.  These plots have been background subtracted by taking the
average intensity in the wedge at $\Theta_{\mathrm{nadir}} =
80^{\circ}$ (beyond the Earth's disk) as a background estimation.  For
comparison the dashed line shows the LAT PSF (averaged between front
and back-converting events) calculated for the lower bound of the
energy bin (the low energy photons dominate the emission in each of
the energy bins).  Note that for the two higher-energy bins the range
and the bin size of these histograms have been reduced to feature
the limb emission.  As can be seen, the Earth emission is wider than
the LAT PSF even for the highest band, in which the LAT PSF is better
than $0.1^{\circ}$ (68\% containment radius). The good agreement in
intensity from north and south can be seen in this figure which
provides a systematic check on the true nature of the east-west
effect.

The angular distribution of photons near the rim can be used to
compare the \gray{} emission with a simple atmospheric model to
investigate the effect of the absorption of \gray{s}.  The proton-air
interaction length is $X_i\sim 85$ g cm$^{-2}$, which can be estimated
using the value $\sim$275 mb for the total inelastic p$^{14}$N cross
section \citep{1996PhRvC..54.1329W}.  In the thin target regime, when
the atmospheric depth is $X\ll X_i$, the intensity of produced
\gray{s} is $\propto X$.  A competing process is the attenuation of
\gray{s} due to the $e^+e^-$-pair production, with the corresponding
scale, the radiation length, of $X_0\sim 38$ g cm$^{-2}$ in the
atmosphere \citep{PDG}.  The interplay between these processes sets
the value of column density when the effect of absorption of \gray{s}
becomes noticeable $\gtrsim 10$ g cm$^{-2}$.

\begin{figure}
  \centering
  \includegraphics[width=\textwidth]{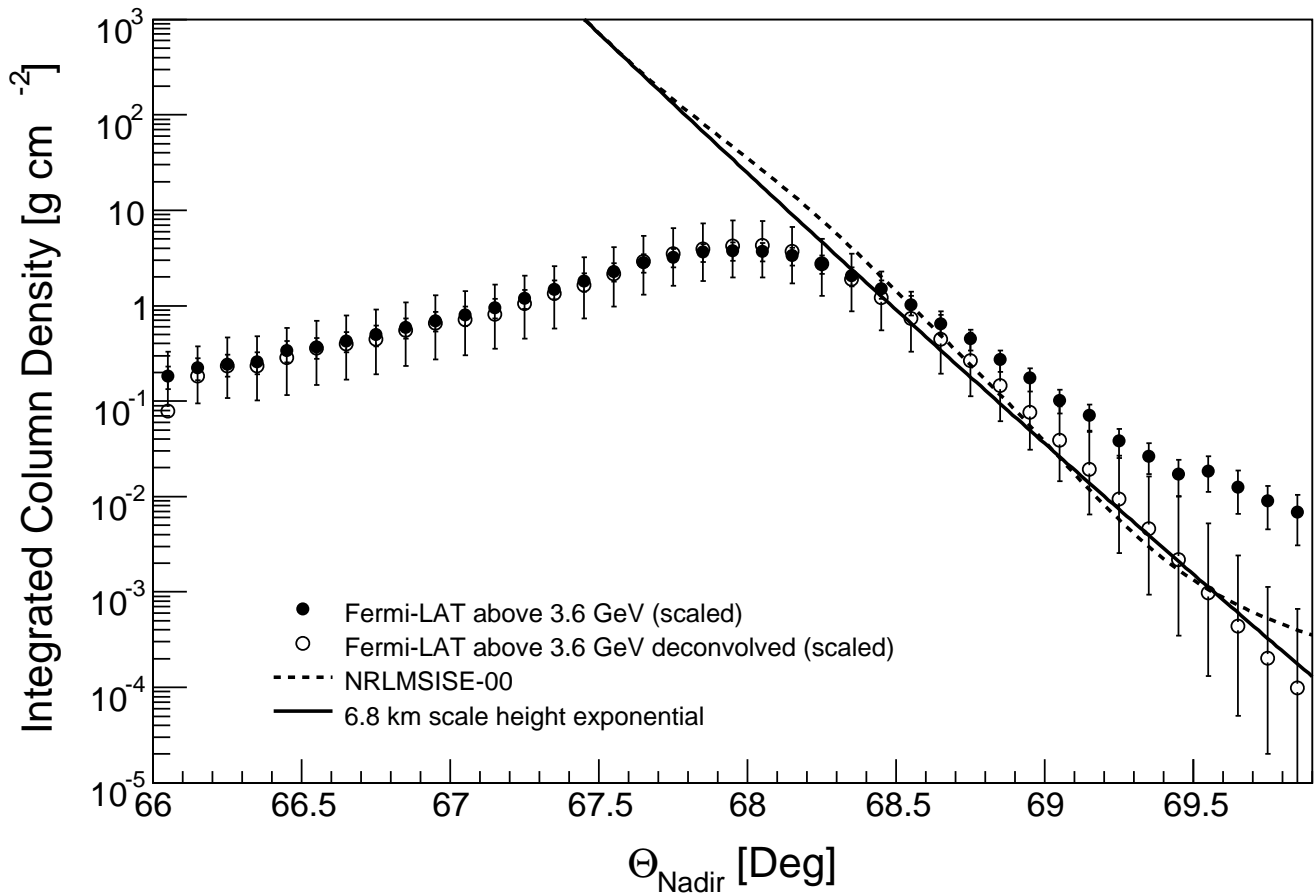}
  \caption{Comparison between the integrated column density with the
    LAT-detected gamma-ray intensity as a function of nadir angle for
    events above 3.6~GeV. Two different model atmospheric profiles have been used: a
    simple barometric atmosphere with scale height of
    6.8~km~\citep{GaisserBook} shown as solid line and a more
    realistic atmospheric model from~\citep{NRLMSISE}. Shown are the
    measured gamma-ray intensity (solid circles) and the gamma-ray intensity
    deconvolved by the PSF of the median energy (4.7~GeV) of this data
    set (open circles).  A good correspondence between the \gray{}
    intensity and the column density can be seen at angles
    $\Theta_{\mathrm{Nadir}} \gtrsim 68.3^{\circ}$. For these angles
    the atmosphere is thin enough (column density $< 10$ g cm$^{-2}$)
    so that no significant attenuation occurs and the \gray{s} are
    directly related to the amount of target material for the incoming
    cosmic rays. The absolute level of the \gray{} intensity has been
    scaled to match the column density for the barometric atmosphere
    model.
    \label{fig::atmosphere}}
\end{figure}

Indeed, this behavior is demonstrated in Figure~\ref{fig::atmosphere},
which shows the integrated atmospheric column depth along the line of
sight for a given nadir angle at a height 560~km for two different
atmospheric models. The solid line shows the curve for a simple model
of the atmosphere with an exponential density profile with scale
height 6.8~km~\citep{GaisserBook} following the {\emph{barometer
    formula}} $X(h) = X(h=0)e^{-(h/h_s)}$ with $X(h)$ in units of g
cm$^{-2}$ denoting the column density of air overlaying a point at
altitude $h$ in cm. The dashed line shows a more realistic atmospheric
model (NRLMSISE-00) as given by~\citet{NRLMSISE}.  Overlaid on the
atmospheric column densities is the \gray{} intensity (scaled to match
the barometric atmosphere profile) averaged over all theta angles
(solid circles). Also shown is the deconvolved LAT intensity (open
circles) in these angles for the PSF at the median energy of this data
set (4.7~GeV). The plot shows, that far enough out beyond the rim (in
the plot for angles $\Theta_{\mathrm{nadir}} > 68.3^{\circ}$), the
\gray{} intensity follows closely the column density. This region is
the {\emph{thin-target}} regime, in which the attenuation of \gray{s}
is not important yet and the intensity scales directly with the amount
of atmosphere for the proton-air interaction. For column densities
larger than $\sim 10$ g cm$^{-2}$ the atmosphere starts to be
optically thick for \gray{s} and absorption (i.e.\ a deviation from
the scaling with density) sets in.

If the target is thin enough that the secondary \gray{s} are
practically not attenuated, the observed \gray{} intensity should have
a power-law index close to the index of ambient cosmic rays. Indeed,
it has been shown that the assumption of a constant fraction
$\kappa\sim 0.17$ of energy of the incident proton released in the
secondary \gray{s} ($\delta$- function approximation) works well in
proton-proton interactions for 1 GeV $\lesssim E_\gamma \lesssim$ 100
GeV \citep{Aharonian2000,Kelner:2006p218}.  The Earth's atmosphere
consists predominantly of N and O, and there is a significant fraction
of He in cosmic rays. This difference in the beam and target
composition could be taken into account by introducing an approximate
energy independent correction factor, but we are currently interested
in the spectral slope and not in the absolute normalization of the
\gray{} intensity. Therefore, a comparison of the ambient spectrum of
protons, with the energy scaled by a factor of $\kappa$, with the
observed spectrum of \gray{s} may be used to check the equality of
their spectral indices and to test the energy calibration of the LAT
instrument.

\begin{figure}
  \centering
  \includegraphics[width=\textwidth]{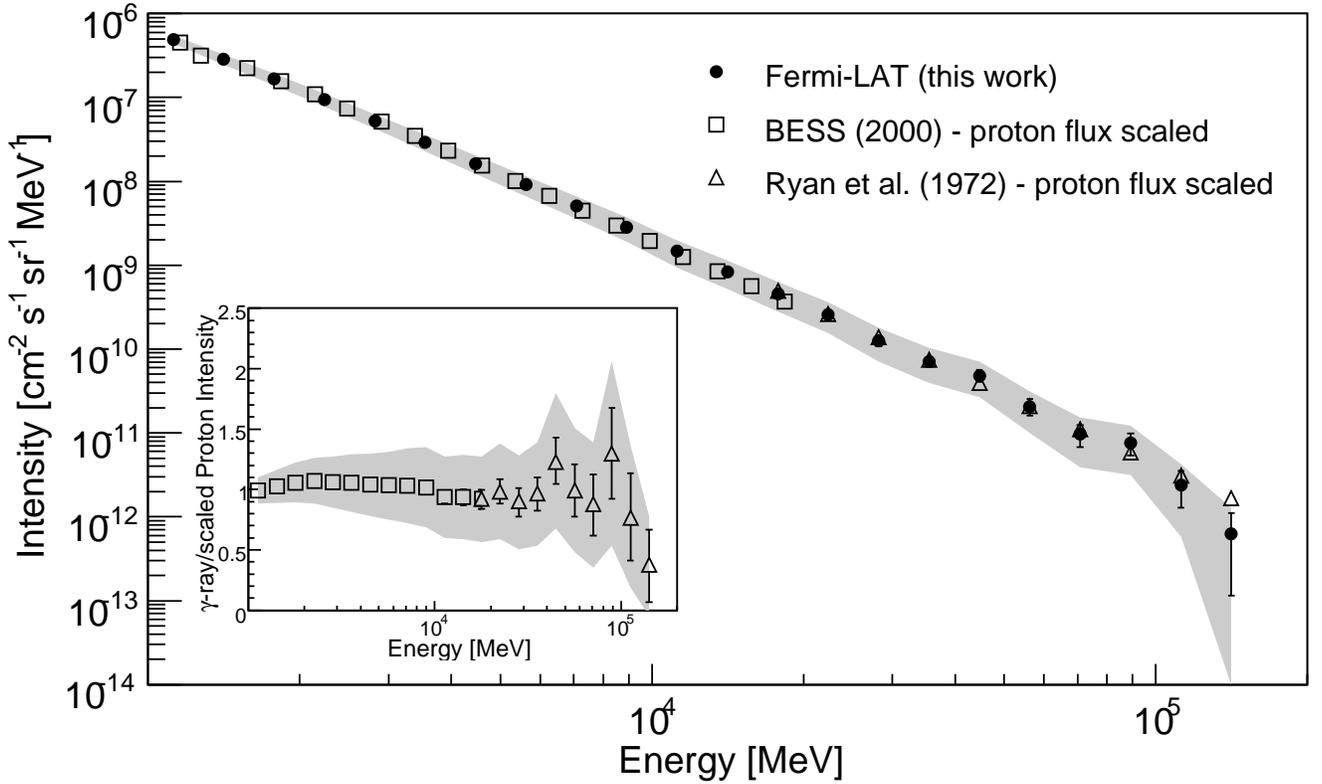}
  \caption{A comparison of the measured gamma-ray intensity for the angular
    interval $68.6^{\circ} < \Theta_{\mathrm{nadir}}< 69.6^{\circ}$
    (thin target case, gamma-ray attenuation is insignificant) with
    the scaled cosmic ray proton intensity (see text for details). The
    inset shows the ratio between the \gray{} data and the 
    scaled proton data and shows the general good match between the
    two measurements. The data have been taken from~\citet{BESS} and
    \citet{ryan1972}. 
    \label{fig::GammaProtonRatio}}
\end{figure}

Figure~\ref{fig::GammaProtonRatio} shows the \gray{} spectrum for the
optically thin region ($68.6^{\circ} < \Theta_{\mathrm{nadir}} <
69.6^{\circ}$) compared to the proton spectrum with the energy scaled
by a factor of $\kappa$. The proton intensity is scaled by first
scaling the energy of each intensity point by a factor $\kappa$, and
then re-normalizing the resulting proton intensity down by 0.7.  In
this way the proton intensity matches the \gray{} intensity at 1~GeV
(i.e., the ratio is normalized to unity at 1~GeV). The general good
agreement is further demonstrated by the inset in
Figure~\ref{fig::GammaProtonRatio} which shows the ratio of the
\gray{} intensity to the scaled proton intensity.

\begin{figure}
  \centering
  \includegraphics[width=\textwidth]{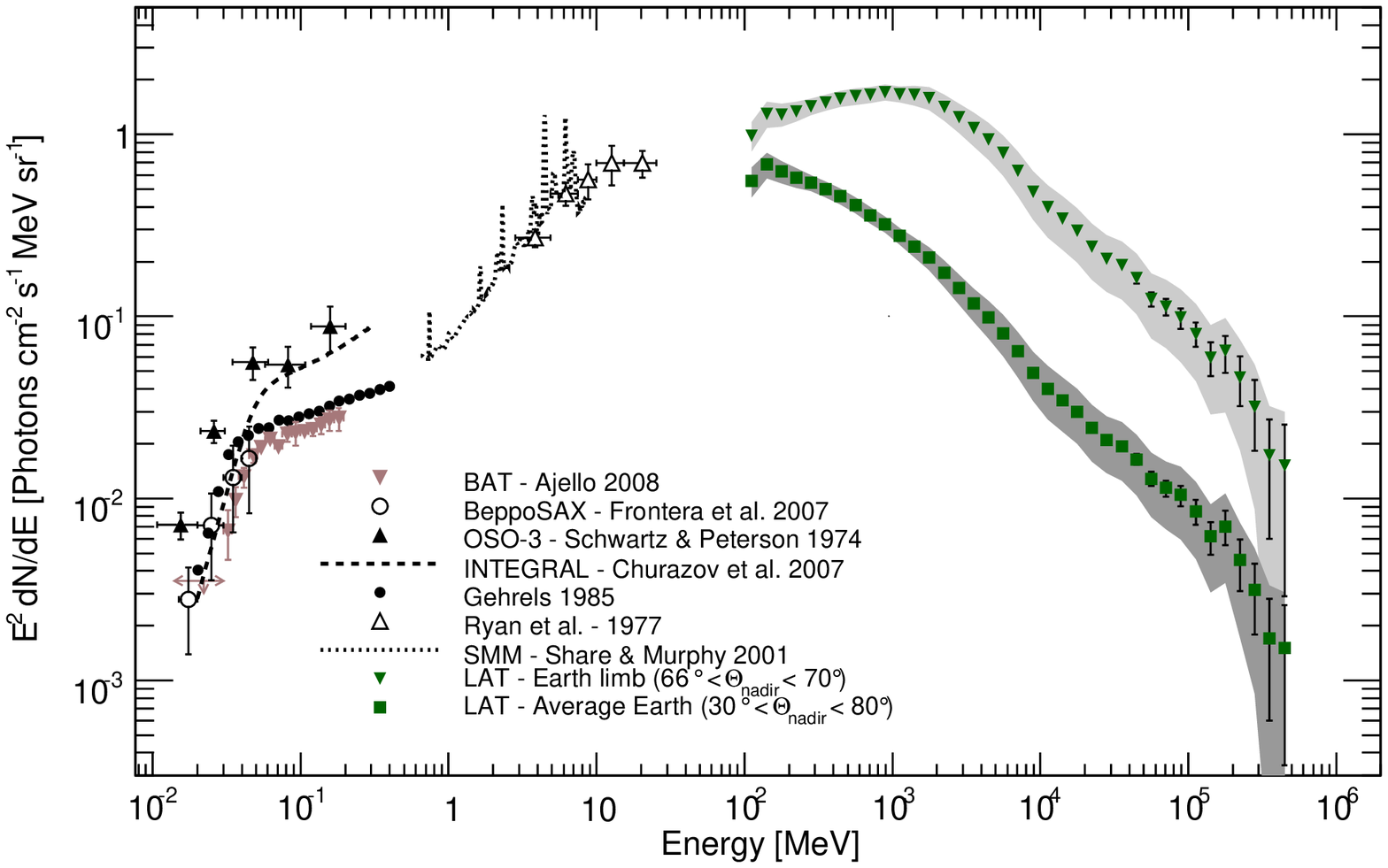}
  \caption{Energy intensity for the Earth-originating \gray{s} from keV to 500
    GeV energies. The measurements from different instruments are not
    readily comparable since they have been taken in different points
    in the solar cycle and are also integrated over different areas
    of the Earth. Data points are taken from~\citet{Ajello2008,
      Frontera2007, Schwartz, Churazov, Gehrels1985, ryan1977, Share2001}.
    \label{fig::SED}}
\end{figure}

\section{Conclusion}
Data from the LAT have been used to study the \gray{} emission
generated in the interactions of cosmic rays with the Earth's
atmosphere for over more than four orders of magnitude between 100~MeV
and 1~TeV with unprecedented precision. The data set contains 218
photons above 100 GeV, and 16 photons above 500 GeV. These
measurements therefore demonstrate the capability of the LAT to detect
and determine energies for photons up to TeV energies.
Two-dimensional intensity maps as well as azimuthal and zenithal
profiles and energy spectra of the Earth albedo emission have been
derived which show several effects:

\begin{itemize}
\item A bright limb at the Earth's horizon.  These limb \gray{s} are
  dominantly generated by grazing-incidence cosmic-ray showers coming
  directly towards the LAT.  The \gray{} spectral index follows the
  spectral index of the incident cosmic ray spectrum in the limb
  region up to \gray{} energies of $E \gg 100 $~GeV.
    
\item A soft-spectrum nadir region dominated by \gray{s}
  back-scattered at large angles originating from cosmic-ray showers
  developing deep into the atmosphere. Due to increasing collimation of
  secondary particles with increasing energies, these \gray{s} show
  a much softer spectrum.
  
\item An east-west modulation for energies below $\sim 10$~GeV, caused
  by the deflection of primary cosmic rays in Earth's magnetic
  field. Higher energy cosmic rays are less affected by the magnetic
  field, resulting in a fading modulation with increasing energy. The
  north-south ratio is equal to 1 within errors.
  
\end{itemize}

Figure~\ref{fig::SED} shows a compilation of photons generated in the
atmosphere of the Earth in the energy range between 10~keV and
1~TeV. A direct comparison of the measurements from different
instruments is not straightforward, since different instruments are
integrating over different regions of the Earth. Also, the
measurements have been taken at different times during the solar cycle
and therefore correspond to different levels of solar modulation
of the primary cosmic rays. The peak of the Earth \gray{} albedo
energy flux is in the LAT range (for the limb, but potentially also
when averaging over the whole Earth). The position of this peak can be
understood from the power-law index of the cosmic-ray spectrum and the
fact that neutral pion production is the dominant process. It is
similarly visible in the spectrum of the diffuse Galactic
emission~\citep{EGRETDiffuse}. Evident in Figure~\ref{fig::SED}
is the huge energy range over which the LAT can make spectral
measurements.

The LAT data provide a picture of the \gray{} emission from the
Earth's atmosphere unprecedented in energy range, resolution and
statistical precision.  The data can be used to understand and model
the interaction of cosmic rays in Earth's atmosphere and magnetic
field. Since the Earth is such a bright source for the LAT, these data
also provide valuable information for \gray{} background
studies and can eventually be used as a calibration source.\\

\acknowledgements{
  The {\textit{Fermi}} LAT Collaboration acknowledges generous ongoing support
  from a number of agencies and institutes that have supported both the
  development and the operation of the LAT as well as scientific data analysis.
  These include the National Aeronautics and Space Administration and the
  Department of Energy in the United States, the Commissariat \`a l'Energie Atomique
  and the Centre National de la Recherche Scientifique / Institut National de Physique
  Nucl\'eaire et de Physique des Particules in France, the Agenzia Spaziale Italiana
  and the Istituto Nazionale di Fisica Nucleare in Italy, the Ministry of Education,
  Culture, Sports, Science and Technology (MEXT), High Energy Accelerator Research
  Organization (KEK) and Japan Aerospace Exploration Agency (JAXA) in Japan, and
  the K.~A.~Wallenberg Foundation, the Swedish Research Council and the
  Swedish National Space Board in Sweden.
  
  Additional support for science analysis during the operations phase is gratefully
  acknowledged from the Istituto Nazionale di Astrofisica in Italy and
  the Centre National d'\'Etudes Spatiales in France. }



\begin{thebibliography}{30}
\expandafter\ifx\csname natexlab\endcsname\relax\def\natexlab#1{#1}\fi
\expandafter\ifx\csname bibnamefont\endcsname\relax
  \def\bibnamefont#1{#1}\fi
\expandafter\ifx\csname bibfnamefont\endcsname\relax
  \def\bibfnamefont#1{#1}\fi
\expandafter\ifx\csname citenamefont\endcsname\relax
  \def\citenamefont#1{#1}\fi
\expandafter\ifx\csname url\endcsname\relax
  \def\url#1{\texttt{#1}}\fi
\expandafter\ifx\csname urlprefix\endcsname\relax\def\urlprefix{URL }\fi
\providecommand{\bibinfo}[2]{#2}
\providecommand{\eprint}[2][]{\url{#2}}

\bibitem[{\citenamefont{{Morris}}(1984)}]{Morris1984}
\bibinfo{author}{\bibfnamefont{D.~J.} \bibnamefont{{Morris}}},
  \bibinfo{journal}{J. Geophys. Res.} \textbf{\bibinfo{volume}{89}},
  \bibinfo{pages}{10685} (\bibinfo{year}{1984}).

\bibitem[{\citenamefont{{Stecker}}(1973)}]{Stecker1973}
\bibinfo{author}{\bibfnamefont{F.~W.} \bibnamefont{{Stecker}}},
  \bibinfo{journal}{\nat} \textbf{\bibinfo{volume}{242}}, \bibinfo{pages}{59}
  (\bibinfo{year}{1973}).

\bibitem[{\citenamefont{{Kraushaar} et~al.}(1972)\citenamefont{{Kraushaar},
  {Clark}, {Garmire}, {Borken}, {Higbie}, {Leong}, and {Thorsos}}}]{OSO3Earth}
\bibinfo{author}{\bibfnamefont{W.~L.} \bibnamefont{{Kraushaar}}},
  \bibinfo{author}{\bibfnamefont{G.~W.} \bibnamefont{{Clark}}},
  \bibinfo{author}{\bibfnamefont{G.~P.} \bibnamefont{{Garmire}}},
  \bibinfo{author}{\bibfnamefont{R.}~\bibnamefont{{Borken}}},
  \bibinfo{author}{\bibfnamefont{P.}~\bibnamefont{{Higbie}}},
  \bibinfo{author}{\bibfnamefont{V.}~\bibnamefont{{Leong}}}, \bibnamefont{and}
  \bibinfo{author}{\bibfnamefont{T.}~\bibnamefont{{Thorsos}}},
  \bibinfo{journal}{\apj} \textbf{\bibinfo{volume}{177}}, \bibinfo{pages}{341}
  (\bibinfo{year}{1972}).

\bibitem[{\citenamefont{{Fichtel} et~al.}(1969)\citenamefont{{Fichtel},
  {Kniffen}, and {{\"O}gelman}}}]{fichtel1969}
\bibinfo{author}{\bibfnamefont{C.~E.} \bibnamefont{{Fichtel}}},
  \bibinfo{author}{\bibfnamefont{D.~A.} \bibnamefont{{Kniffen}}},
  \bibnamefont{and} \bibinfo{author}{\bibfnamefont{H.~B.}
  \bibnamefont{{{\"O}gelman}}}, \bibinfo{journal}{\apj}
  \textbf{\bibinfo{volume}{158}}, \bibinfo{pages}{193} (\bibinfo{year}{1969}).

\bibitem[{\citenamefont{{Kinzer} et~al.}(1974)\citenamefont{{Kinzer}, {Share},
  and {Seeman}}}]{kinzer1974}
\bibinfo{author}{\bibfnamefont{R.~L.} \bibnamefont{{Kinzer}}},
  \bibinfo{author}{\bibfnamefont{G.~H.} \bibnamefont{{Share}}},
  \bibnamefont{and} \bibinfo{author}{\bibfnamefont{N.}~\bibnamefont{{Seeman}}},
  \bibinfo{journal}{J. Geophys. Res.} \textbf{\bibinfo{volume}{79}},
  \bibinfo{pages}{4567} (\bibinfo{year}{1974}).

\bibitem[{\citenamefont{{Fishman} et~al.}(1976)\citenamefont{{Fishman},
  {Watts}, and {Meegan}}}]{fishman1976}
\bibinfo{author}{\bibfnamefont{G.~J.} \bibnamefont{{Fishman}}},
  \bibinfo{author}{\bibfnamefont{J.~W.} \bibnamefont{{Watts}},
  \bibfnamefont{Jr.}}, \bibnamefont{and} \bibinfo{author}{\bibfnamefont{C.~A.}
  \bibnamefont{{Meegan}}}, \bibinfo{journal}{J. Geophys. Res.}
  \textbf{\bibinfo{volume}{81}}, \bibinfo{pages}{6121} (\bibinfo{year}{1976}).

\bibitem[{\citenamefont{{Ryan} et~al.}(1977)\citenamefont{{Ryan}, {Moon},
  {Wilson}, {Zych}, {White}, and {Dayton}}}]{ryan1977}
\bibinfo{author}{\bibfnamefont{J.~M.} \bibnamefont{{Ryan}}},
  \bibinfo{author}{\bibfnamefont{S.~H.} \bibnamefont{{Moon}}},
  \bibinfo{author}{\bibfnamefont{R.~B.} \bibnamefont{{Wilson}}},
  \bibinfo{author}{\bibfnamefont{A.~D.} \bibnamefont{{Zych}}},
  \bibinfo{author}{\bibfnamefont{R.~S.} \bibnamefont{{White}}},
  \bibnamefont{and} \bibinfo{author}{\bibfnamefont{B.}~\bibnamefont{{Dayton}}},
  \bibinfo{journal}{J. Geophys. Res.} \textbf{\bibinfo{volume}{82}},
  \bibinfo{pages}{3593} (\bibinfo{year}{1977}).

\bibitem[{\citenamefont{{Schoenfelder}
  et~al.}(1977)\citenamefont{{Schoenfelder}, {Graser}, and
  {Daugherty}}}]{schoenfelder1977}
\bibinfo{author}{\bibfnamefont{V.}~\bibnamefont{{Schoenfelder}}},
  \bibinfo{author}{\bibfnamefont{U.}~\bibnamefont{{Graser}}}, \bibnamefont{and}
  \bibinfo{author}{\bibfnamefont{J.}~\bibnamefont{{Daugherty}}},
  \bibinfo{journal}{\apj} \textbf{\bibinfo{volume}{217}}, \bibinfo{pages}{306}
  (\bibinfo{year}{1977}).

\bibitem[{\citenamefont{{Golenetskii} et~al.}(1975)\citenamefont{{Golenetskii},
  {Gurian}, {Ilinskii}, {Mazetz}, and {Proskura}}}]{golenetskiy1975}
\bibinfo{author}{\bibfnamefont{S.~V.} \bibnamefont{{Golenetskii}}},
  \bibinfo{author}{\bibfnamefont{I.~A.} \bibnamefont{{Gurian}}},
  \bibinfo{author}{\bibfnamefont{V.~N.} \bibnamefont{{Ilinskii}}},
  \bibinfo{author}{\bibfnamefont{E.~P.} \bibnamefont{{Mazetz}}},
  \bibnamefont{and} \bibinfo{author}{\bibfnamefont{M.~P.}
  \bibnamefont{{Proskura}}}, \bibinfo{journal}{Geomagnetism and
  Aeronomy/Geomagnetizm i Aeronomiia} \textbf{\bibinfo{volume}{15}},
  \bibinfo{pages}{203} (\bibinfo{year}{1975}).

\bibitem[{\citenamefont{{Imhof} et~al.}(1976)\citenamefont{{Imhof}, {Nakano},
  and {Reagan}}}]{imhof1976}
\bibinfo{author}{\bibfnamefont{W.~L.} \bibnamefont{{Imhof}}},
  \bibinfo{author}{\bibfnamefont{G.~H.} \bibnamefont{{Nakano}}},
  \bibnamefont{and} \bibinfo{author}{\bibfnamefont{J.~B.}
  \bibnamefont{{Reagan}}}, \bibinfo{journal}{J. Geophys. Res.}
  \textbf{\bibinfo{volume}{81}}, \bibinfo{pages}{2835} (\bibinfo{year}{1976}).

\bibitem[{\citenamefont{{Gurian} et~al.}(1979)\citenamefont{{Gurian}, {Mazets},
  {Proskura}, and {Sokolov}}}]{guryan1979}
\bibinfo{author}{\bibfnamefont{I.~A.} \bibnamefont{{Gurian}}},
  \bibinfo{author}{\bibfnamefont{E.~P.} \bibnamefont{{Mazets}}},
  \bibinfo{author}{\bibfnamefont{M.~P.} \bibnamefont{{Proskura}}},
  \bibnamefont{and} \bibinfo{author}{\bibfnamefont{I.~A.}
  \bibnamefont{{Sokolov}}}, \bibinfo{journal}{Geomagnetism and
  Aeronomy/Geomagnetizm i Aeronomiia} \textbf{\bibinfo{volume}{19}},
  \bibinfo{pages}{11} (\bibinfo{year}{1979}).

\bibitem[{\citenamefont{{Share} and {Murphy}}(2001)}]{Share2001}
\bibinfo{author}{\bibfnamefont{G.~H.} \bibnamefont{{Share}}} \bibnamefont{and}
  \bibinfo{author}{\bibfnamefont{R.~J.} \bibnamefont{{Murphy}}},
  \bibinfo{journal}{J. Geophys. Res.} \textbf{\bibinfo{volume}{106}},
  \bibinfo{pages}{77} (\bibinfo{year}{2001}).

\bibitem[{\citenamefont{{Thompson} et~al.}(1981)\citenamefont{{Thompson},
  {Simpson}, and {Ozel}}}]{SASEarth}
\bibinfo{author}{\bibfnamefont{D.~J.} \bibnamefont{{Thompson}}},
  \bibinfo{author}{\bibfnamefont{G.~A.} \bibnamefont{{Simpson}}},
  \bibnamefont{and} \bibinfo{author}{\bibfnamefont{M.~E.}
  \bibnamefont{{Ozel}}}, \bibinfo{journal}{J. Geophys. Res.}
  \textbf{\bibinfo{volume}{86}}, \bibinfo{pages}{1265} (\bibinfo{year}{1981}).

\bibitem[{\citenamefont{{Petry}}(2005)}]{PetryEarth}
\bibinfo{author}{\bibfnamefont{D.}~\bibnamefont{{Petry}}}, in
  \emph{\bibinfo{booktitle}{High Energy Gamma-Ray Astronomy}}, edited by
  \bibinfo{editor}{\bibfnamefont{F.~A.} \bibnamefont{{Aharonian}}},
  \bibinfo{editor}{\bibfnamefont{H.~J.} \bibnamefont{{V{\"o}lk}}},
  \bibnamefont{and} \bibinfo{editor}{\bibfnamefont{D.}~\bibnamefont{{Horns}}}
  (\bibinfo{year}{2005}), vol. \bibinfo{volume}{745} of
  \emph{\bibinfo{series}{American Institute of Physics Conference Series}}, pp.
  \bibinfo{pages}{709--714}.

\bibitem[{\citenamefont{{Fermi/LAT Collaboration:
  W.~B.~Atwood}}(2009)}]{LATPaper}
\bibinfo{author}{\bibnamefont{{Fermi/LAT Collaboration: W.~B.~Atwood}}},
  \bibinfo{journal}{\apj} \textbf{\bibinfo{volume}{697}}, \bibinfo{pages}{1071}
  (\bibinfo{year}{2009}), \eprint{0902.1089}.

\bibitem[{\citenamefont{{Abdo} et~al.}(2009)\citenamefont{{Abdo}, {Ackermann},
  {Atwood}, {Bagagli}, {Baldini}, {Ballet}, {Band}, {Barbiellini}, {Baring},
  {Bartelt} et~al.}}]{FermiVela}
\bibinfo{author}{\bibfnamefont{A.~A.} \bibnamefont{{Abdo}}},
  \bibinfo{author}{\bibfnamefont{M.}~\bibnamefont{{Ackermann}}},
  \bibinfo{author}{\bibfnamefont{W.~B.} \bibnamefont{{Atwood}}},
  \bibinfo{author}{\bibfnamefont{R.}~\bibnamefont{{Bagagli}}},
  \bibinfo{author}{\bibfnamefont{L.}~\bibnamefont{{Baldini}}},
  \bibinfo{author}{\bibfnamefont{J.}~\bibnamefont{{Ballet}}},
  \bibinfo{author}{\bibfnamefont{D.~L.} \bibnamefont{{Band}}},
  \bibinfo{author}{\bibfnamefont{G.}~\bibnamefont{{Barbiellini}}},
  \bibinfo{author}{\bibfnamefont{M.~G.} \bibnamefont{{Baring}}},
  \bibinfo{author}{\bibfnamefont{J.}~\bibnamefont{{Bartelt}}},
  \bibnamefont{et~al.}, \bibinfo{journal}{\apj} \textbf{\bibinfo{volume}{696}},
  \bibinfo{pages}{1084} (\bibinfo{year}{2009}), \eprint{0812.2960}.

\bibitem[{\citenamefont{{Ryan} et~al.}(1972)\citenamefont{{Ryan}, {Ormes}, and
  {Balasubrahmanyan}}}]{ryan1972}
\bibinfo{author}{\bibfnamefont{M.~J.} \bibnamefont{{Ryan}}},
  \bibinfo{author}{\bibfnamefont{J.~F.} \bibnamefont{{Ormes}}},
  \bibnamefont{and} \bibinfo{author}{\bibfnamefont{V.~K.}
  \bibnamefont{{Balasubrahmanyan}}}, \bibinfo{journal}{Physical Review Letters}
  \textbf{\bibinfo{volume}{28}}, \bibinfo{pages}{985} (\bibinfo{year}{1972}).

\bibitem[{\citenamefont{{Sanuki} et~al.}(2000)\citenamefont{{Sanuki}, {Motoki},
  {Matsumoto}, {Seo}, {Wang}, {Abe}, {Anraku}, {Asaoka}, {Fujikawa}, {Imori}
  et~al.}}]{BESS}
\bibinfo{author}{\bibfnamefont{T.}~\bibnamefont{{Sanuki}}},
  \bibinfo{author}{\bibfnamefont{M.}~\bibnamefont{{Motoki}}},
  \bibinfo{author}{\bibfnamefont{H.}~\bibnamefont{{Matsumoto}}},
  \bibinfo{author}{\bibfnamefont{E.~S.} \bibnamefont{{Seo}}},
  \bibinfo{author}{\bibfnamefont{J.~Z.} \bibnamefont{{Wang}}},
  \bibinfo{author}{\bibfnamefont{K.}~\bibnamefont{{Abe}}},
  \bibinfo{author}{\bibfnamefont{K.}~\bibnamefont{{Anraku}}},
  \bibinfo{author}{\bibfnamefont{Y.}~\bibnamefont{{Asaoka}}},
  \bibinfo{author}{\bibfnamefont{M.}~\bibnamefont{{Fujikawa}}},
  \bibinfo{author}{\bibfnamefont{M.}~\bibnamefont{{Imori}}},
  \bibnamefont{et~al.}, \bibinfo{journal}{\apj} \textbf{\bibinfo{volume}{545}},
  \bibinfo{pages}{1135} (\bibinfo{year}{2000}),
  \eprint{arXiv:astro-ph/0002481}.

\bibitem[{\citenamefont{{Aharonian} and {Atoyan}}(2000)}]{Aharonian2000}
\bibinfo{author}{\bibfnamefont{F.~A.} \bibnamefont{{Aharonian}}}
  \bibnamefont{and} \bibinfo{author}{\bibfnamefont{A.~M.}
  \bibnamefont{{Atoyan}}}, \bibinfo{journal}{Astron. Astrophys.}
  \textbf{\bibinfo{volume}{362}}, \bibinfo{pages}{937} (\bibinfo{year}{2000}),
  \eprint{arXiv:astro-ph/0009009}.

\bibitem[{\citenamefont{{Wellisch} and {Axen}}(1996)}]{1996PhRvC..54.1329W}
\bibinfo{author}{\bibfnamefont{H.~P.} \bibnamefont{{Wellisch}}}
  \bibnamefont{and} \bibinfo{author}{\bibfnamefont{D.}~\bibnamefont{{Axen}}},
  \bibinfo{journal}{\prc} \textbf{\bibinfo{volume}{54}}, \bibinfo{pages}{1329}
  (\bibinfo{year}{1996}).

\bibitem[{\citenamefont{Amsler et~al.}(2008)\citenamefont{Amsler, Doser,
  Antonelli, Asner, Babu, Baer, Band, Barnett, Bergren, Beringer et~al.}}]{PDG}
\bibinfo{author}{\bibfnamefont{C.}~\bibnamefont{Amsler}},
  \bibinfo{author}{\bibfnamefont{M.}~\bibnamefont{Doser}},
  \bibinfo{author}{\bibfnamefont{M.}~\bibnamefont{Antonelli}},
  \bibinfo{author}{\bibfnamefont{D.}~\bibnamefont{Asner}},
  \bibinfo{author}{\bibfnamefont{K.}~\bibnamefont{Babu}},
  \bibinfo{author}{\bibfnamefont{H.}~\bibnamefont{Baer}},
  \bibinfo{author}{\bibfnamefont{H.}~\bibnamefont{Band}},
  \bibinfo{author}{\bibfnamefont{R.}~\bibnamefont{Barnett}},
  \bibinfo{author}{\bibfnamefont{E.}~\bibnamefont{Bergren}},
  \bibinfo{author}{\bibfnamefont{J.}~\bibnamefont{Beringer}},
  \bibnamefont{et~al.}, \bibinfo{journal}{Physics Letters B}
  \textbf{\bibinfo{volume}{667}}, \bibinfo{pages}{1 } (\bibinfo{year}{2008}),
  ISSN \bibinfo{issn}{0370-2693}, \bibinfo{note}{review of Particle Physics}.

\bibitem[{\citenamefont{Gaisser}(1990)}]{GaisserBook}
\bibinfo{author}{\bibfnamefont{T.~K.} \bibnamefont{Gaisser}},
  \emph{\bibinfo{title}{Cosmic rays and particle physics}}
  (\bibinfo{publisher}{Cambridge Univ. Press}, \bibinfo{address}{Cambridge},
  \bibinfo{year}{1990}).

\bibitem[{\citenamefont{{Picone} et~al.}(2002)\citenamefont{{Picone}, {Hedin},
  {Drob}, and {Aikin}}}]{NRLMSISE}
\bibinfo{author}{\bibfnamefont{J.~M.} \bibnamefont{{Picone}}},
  \bibinfo{author}{\bibfnamefont{A.~E.} \bibnamefont{{Hedin}}},
  \bibinfo{author}{\bibfnamefont{D.~P.} \bibnamefont{{Drob}}},
  \bibnamefont{and} \bibinfo{author}{\bibfnamefont{A.~C.}
  \bibnamefont{{Aikin}}}, \bibinfo{journal}{Journal of Geophysical Research
  (Space Physics)} \textbf{\bibinfo{volume}{107}}, \bibinfo{pages}{1468}
  (\bibinfo{year}{2002}).

\bibitem[{\citenamefont{Kelner et~al.}(2006)\citenamefont{Kelner, Aharonian,
  and Bugayov}}]{Kelner:2006p218}
\bibinfo{author}{\bibfnamefont{S.~R.} \bibnamefont{Kelner}},
  \bibinfo{author}{\bibfnamefont{F.~A.} \bibnamefont{Aharonian}},
  \bibnamefont{and} \bibinfo{author}{\bibfnamefont{V.~V.}
  \bibnamefont{Bugayov}}, \bibinfo{journal}{Phys. Rev. D}
  \textbf{\bibinfo{volume}{74}}, \bibinfo{pages}{16} (\bibinfo{year}{2006}).

\bibitem[{\citenamefont{{Ajello} et~al.}(2008)\citenamefont{{Ajello},
  {Greiner}, {Sato}, {Willis}, {Kanbach}, {Strong}, {Diehl}, {Hasinger},
  {Gehrels}, {Markwardt} et~al.}}]{Ajello2008}
\bibinfo{author}{\bibfnamefont{M.}~\bibnamefont{{Ajello}}},
  \bibinfo{author}{\bibfnamefont{J.}~\bibnamefont{{Greiner}}},
  \bibinfo{author}{\bibfnamefont{G.}~\bibnamefont{{Sato}}},
  \bibinfo{author}{\bibfnamefont{D.~R.} \bibnamefont{{Willis}}},
  \bibinfo{author}{\bibfnamefont{G.}~\bibnamefont{{Kanbach}}},
  \bibinfo{author}{\bibfnamefont{A.~W.} \bibnamefont{{Strong}}},
  \bibinfo{author}{\bibfnamefont{R.}~\bibnamefont{{Diehl}}},
  \bibinfo{author}{\bibfnamefont{G.}~\bibnamefont{{Hasinger}}},
  \bibinfo{author}{\bibfnamefont{N.}~\bibnamefont{{Gehrels}}},
  \bibinfo{author}{\bibfnamefont{C.~B.} \bibnamefont{{Markwardt}}},
  \bibnamefont{et~al.}, \bibinfo{journal}{\apj} \textbf{\bibinfo{volume}{689}},
  \bibinfo{pages}{666} (\bibinfo{year}{2008}), \eprint{0808.3377}.

\bibitem[{\citenamefont{{Frontera} et~al.}(2007)\citenamefont{{Frontera},
  {Orlandini}, {Landi}, {Comastri}, {Fiore}, {Setti}, {Amati}, {Costa},
  {Masetti}, and {Palazzi}}}]{Frontera2007}
\bibinfo{author}{\bibfnamefont{F.}~\bibnamefont{{Frontera}}},
  \bibinfo{author}{\bibfnamefont{M.}~\bibnamefont{{Orlandini}}},
  \bibinfo{author}{\bibfnamefont{R.}~\bibnamefont{{Landi}}},
  \bibinfo{author}{\bibfnamefont{A.}~\bibnamefont{{Comastri}}},
  \bibinfo{author}{\bibfnamefont{F.}~\bibnamefont{{Fiore}}},
  \bibinfo{author}{\bibfnamefont{G.}~\bibnamefont{{Setti}}},
  \bibinfo{author}{\bibfnamefont{L.}~\bibnamefont{{Amati}}},
  \bibinfo{author}{\bibfnamefont{E.}~\bibnamefont{{Costa}}},
  \bibinfo{author}{\bibfnamefont{N.}~\bibnamefont{{Masetti}}},
  \bibnamefont{and}
  \bibinfo{author}{\bibfnamefont{E.}~\bibnamefont{{Palazzi}}},
  \bibinfo{journal}{\apj} \textbf{\bibinfo{volume}{666}}, \bibinfo{pages}{86}
  (\bibinfo{year}{2007}), \eprint{arXiv:astro-ph/0611228}.

\bibitem[{\citenamefont{{Schwartz} and {Peterson}}(1974)}]{Schwartz}
\bibinfo{author}{\bibfnamefont{D.~A.} \bibnamefont{{Schwartz}}}
  \bibnamefont{and} \bibinfo{author}{\bibfnamefont{L.~E.}
  \bibnamefont{{Peterson}}}, \bibinfo{journal}{\apj}
  \textbf{\bibinfo{volume}{190}}, \bibinfo{pages}{297} (\bibinfo{year}{1974}).

\bibitem[{\citenamefont{{Churazov} et~al.}(2007)\citenamefont{{Churazov},
  {Sunyaev}, {Revnivtsev}, {Sazonov}, {Molkov}, {Grebenev}, {Winkler},
  {Parmar}, {Bazzano}, {Falanga} et~al.}}]{Churazov}
\bibinfo{author}{\bibfnamefont{E.}~\bibnamefont{{Churazov}}},
  \bibinfo{author}{\bibfnamefont{R.}~\bibnamefont{{Sunyaev}}},
  \bibinfo{author}{\bibfnamefont{M.}~\bibnamefont{{Revnivtsev}}},
  \bibinfo{author}{\bibfnamefont{S.}~\bibnamefont{{Sazonov}}},
  \bibinfo{author}{\bibfnamefont{S.}~\bibnamefont{{Molkov}}},
  \bibinfo{author}{\bibfnamefont{S.}~\bibnamefont{{Grebenev}}},
  \bibinfo{author}{\bibfnamefont{C.}~\bibnamefont{{Winkler}}},
  \bibinfo{author}{\bibfnamefont{A.}~\bibnamefont{{Parmar}}},
  \bibinfo{author}{\bibfnamefont{A.}~\bibnamefont{{Bazzano}}},
  \bibinfo{author}{\bibfnamefont{M.}~\bibnamefont{{Falanga}}},
  \bibnamefont{et~al.}, \bibinfo{journal}{\aa} \textbf{\bibinfo{volume}{467}},
  \bibinfo{pages}{529} (\bibinfo{year}{2007}), \eprint{arXiv:astro-ph/0608250}.

\bibitem[{\citenamefont{{Gehrels}}(1985)}]{Gehrels1985}
\bibinfo{author}{\bibfnamefont{N.}~\bibnamefont{{Gehrels}}},
  \bibinfo{journal}{Nuclear Instruments and Methods in Physics Research A}
  \textbf{\bibinfo{volume}{239}}, \bibinfo{pages}{324} (\bibinfo{year}{1985}).

\bibitem[{\citenamefont{{Strong} et~al.}(2004)\citenamefont{{Strong},
  {Moskalenko}, and {Reimer}}}]{EGRETDiffuse}
\bibinfo{author}{\bibfnamefont{A.~W.} \bibnamefont{{Strong}}},
  \bibinfo{author}{\bibfnamefont{I.~V.} \bibnamefont{{Moskalenko}}},
  \bibnamefont{and} \bibinfo{author}{\bibfnamefont{O.}~\bibnamefont{{Reimer}}},
  \bibinfo{journal}{\apj} \textbf{\bibinfo{volume}{613}}, \bibinfo{pages}{962}
  (\bibinfo{year}{2004}), \eprint{arXiv:astro-ph/0406254}.

\end{thebibliography}

\end{document}